\newcounter{subequation}[equation]
\def\thesubequation{\theequation\@alph\c@subequation}
\def\@subeqnnum{{\rm (\thesubequation)}}
\def\slabel#1{\@bsphack\if@filesw {\let\thepage\relax
   \xdef\@gtempa{\write\@auxout{\string
      \newlabel{#1}{{\thesubequation}{\thepage}}}}}\@gtempa
   \if@nobreak \ifvmode\nobreak\fi\fi\fi\@esphack}
\def\subeqnarray{\stepcounter{equation}
\let\@currentlabel=\theequation\global\c@subequation\@ne
\global\@eqnswtrue
\global\@eqcnt\z@\tabskip\@centering\let\\=\@subeqncr
$$\halign to \displaywidth\bgroup\@eqnsel\hskip\@centering
  $\displaystyle\tabskip\z@{##}$&\global\@eqcnt\@ne
  \hskip 2\arraycolsep \hfil${##}$\hfil
  &\global\@eqcnt\tw@ \hskip 2\arraycolsep $\displaystyle\tabskip\z@{##}$\hfil
   \tabskip\@centering&\llap{##}\tabskip\z@\cr}

\def\endsubeqnarray{\@@subeqncr\egroup
                     $$\global\@ignoretrue}

\def\@subeqncr{{\ifnum0=`}\fi\@ifstar{\global\@eqpen\@M
    \@ysubeqncr}{\global\@eqpen\interdisplaylinepenalty \@ysubeqncr}}

\def\@ysubeqncr{\@ifnextchar [{\@xsubeqncr}{\@xsubeqncr[\z@]}}

\def\@xsubeqncr[#1]{\ifnum0=`{\fi}\@@subeqncr
   \noalign{\penalty\@eqpen\vskip\jot\vskip #1\relax}}

\def\@@subeqncr{\let\@tempa\relax
    \ifcase\@eqcnt \def\@tempa{& & &}\or \def\@tempa{& &}
      \else \def\@tempa{&}\fi
     \@tempa \if@eqnsw\@subeqnnum\refstepcounter{subequation}\fi
     \global\@eqnswtrue\global\@eqcnt\z@\cr}

\let\@ssubeqncr=\@subeqncr
\@namedef{subeqnarray*}{\def\@subeqncr{\nonumber\@ssubeqncr}\subeqnarray}
\@namedef{endsubeqnarray*}{\global\advance\c@equation\m@ne%
                           \nonumber\endsubeqnarray}
\documentstyle[rep12,subeqnar]{article}
\newlength{\dinwidth}
\newlength{\dinmargin}
\def\preprint#1{\gdef\@preprint{#1}}
\def\@prprint{
   \begin{flushright}
     \begin{tabular}{l}
        \@preprint \\
        \@date
     \end{tabular}
   \end{flushright}}
\def\@maketitle{\newpage
\@prprint
 \vskip 3em \begin{center}
 {\LARGE \@title \par} \vskip 1.5em {\large \lineskip .5em
\begin{tabular}[t]{c}\@author
 \end{tabular}\par}
\vskip 1em
                            \end{center}
 \par
 \vskip 3.5em}
\def\be{\begin{equation}}
\def\ee{\end{equation}}
\font\cmss=cmss10 \font\cmsss=cmss10 at 10truept
\def\IZ{\relax\ifmmode\mathchoice
{\hbox{\cmss Z\kern-.4em Z}}{\hbox{\cmss Z\kern-.4em Z}}
{\lower.9pt\hbox{\cmsss Z\kern-.36em Z}}
{\lower1.2pt\hbox{\cmsss Z\kern-.36em Z}}
\else{\cmss Z\kern-.4em Z}\fi}
\def\ii{\'{\i}}
\def\reff#1{(\ref{#1})}
\preprint{FTUAM 93-18 \\ DFTUZ 93.3}
\title{Finite Size Analysis of the One-dimensional
       $q = \infty$ Clock Model}
\author{{\bf M. Asorey}\thanks{e-mail
       asorey@saturno.unizar.es},
       {\bf J.G. Esteve}\thanks{e-mail esteve@jupiter.unizar.es} \\
       {\em Departamento de F\ii sica Te\'orica} \\
       {\em Universidad de Zaragoza}  \\
       {\em 50009 Zaragoza. SPAIN} \\
       and \\
       {\bf J. Salas}\thanks{e-mail duncan@vm1.sdi.uam.es} \\
       {\em Departamento de F\ii sica Te\'orica C-XI} \\
       {\em Universidad Aut\'onoma de Madrid}  \\
       {\em Cantoblanco 28049 Madrid. SPAIN} }
\begin{document}
\maketitle
\thispagestyle{empty}
\begin{abstract}
We analyze the finite size scaling of the $q$-state clock model
in the  $q \rightarrow \infty$ limit. The behaviors of the specific
heat, Binder-Landau and U4 cumulants agree with  the Borgs-Koteck\'y
ans\"atz for  first order phase transitions. However, we find that the
leading correction to the position of the extremal points of these
quantities is not universal. On the other hand, the finite size
corrections to the mass gap  behave like for second order phase
transitions.
In particular, the curves corresponding to different
size approximations do  not cross in  the vicinity of the  transition
points. The feature is associated to the existence of a divergent
correlation length and holds for a wider class of models.
\end{abstract}
\newpage

\section{Introduction}

In the last few years there has been a renewed activity on the
study of first order phase transitions motivated in part by
some controversies over the behavior of the renormalization group
in the vicinity of the transition  and the
criteria to determine the order of a transition from  finite size
analysis. Although there are not many new
results, however, two different rigorous
approaches have  shed some light on  those problems.

For systems with bounded
fluctuating variables and absolutely summable real Hamiltonians it
has been  rigorously proved
that the renormalization group when properly defined is
continuous and single-valued across the transition surface.
The only real pathologies   arise  at some points  where the
renormalization group is not defined at all \cite{sokal}.

On the other hand, a rigorous analysis of the finite size scaling
was developed by Borgs and Koteck\'y for systems which admit a
representation as contour models with small activities
such as the Ising
model at low temperature and the $q$-state Potts model with $q$ large
enough \cite{borgs-kotecky}. In such a case, for large lattices
sizes $L>> 1$ the partition function
can be written as a sum of terms, each of them giving the
contribution of small fluctuations around a pure phase, plus a
remainder, which can be bounded by an exponentially decaying term
${\cal O}(e^{-bL})$. Consequently, the asymptotic behavior
of the extremal points of  the specific heat (CV),
Binder--Landau cumulant (BL) \cite{binder-landau},
\be
BL = {1 \over 3} \left( 1 - \frac{\langle E^4 \rangle}
     {\langle E^2 \rangle^2} \right)
\ee
and  $U4$ ratio of cumulants,
\be
U4 = \frac{\langle (E - \langle E \rangle^4 \rangle}
         {\langle (E - \langle E \rangle)^2 \rangle^2}.
\ee
in a $d$-dimensional system is given in terms of increasing powers
of $L^{-d}$ plus some
exponentially decaying terms. If  $Q(L)$ denote the value of
any of the quantities $CV/L^d$, $U4$ or $BL$ in a square lattice of
size $L$ and $\beta^Q_\ast(L)$ the inverse of the temperatures of the
corresponding extremal points, the Borgs-Koteck\'y results imply that
\begin{subeqnarray}
\slabel{ansatz1}
   Q_\ast(L)        &=& X^{(0)}_Q + X^{(1)}_Q/L^d + {\cal O}(L^{-2d})
\\ \slabel{ansatz2}
   \beta_\ast^Q(L) &=& \beta_c   + Y^{(1)}_Q/L^d + Y^{(2)}_Q/L^{2d} +
                                            {\cal O}(L^{-3d}),
\end{subeqnarray}
where $\beta_c$ is the inverse of the critical temperature.
The precise values of the coefficients  $\{X^{(k)}_Q\}_{k=0,1}$ and
$\{Y^{(k)}_Q\}_{k=1,2}$  are exactly known for  the $q$-state Potts
model for large values $q$ \cite{billoire10}. This behavior has
been numerically verified for the $q=20$ Potts model in two
dimensional lattices \cite{billoire20}. However, in the case of lower
q-state Potts models (q=7,10) Monte Carlo calculations for
square lattices up to size $L= 50$ are not in agreement with this
ans\"atz \cite{billoire10}. The disagreement might be due to the
fact that the lattice size is not large enough to reach the
asymptotic behavior. Otherwise, it would mean that the
Borgs-Koteck\`y ans\"atz does not hold in such a case.
In any case the Borgs-Koteck\`y ans\"atz implies that $\nu=1/d $
in agreement with the value predicted by the discontinuity fixed
point scenario of the renormalization group behavior for first order
transitions \cite{dfp}. Moreover, if $X_{CV/L^d}^{(0)} \neq 0$ the finite
size analysis  becomes similar to
that of second order phase transitions with $\alpha =1$, being
the existence of as
many critical exponents $y=d$ as different phases can coexist the main
signal of the first order  transitions \cite{dfp}.

In this note we analyze  these problems in the
one-dimensional $q$-state clock model introduced by two of us
\cite{asorey-esteve}.
The simplicity of the model implies that it can be exactly solved and the
quantities can be exactly computed  without
numerical errors.

\section{Finite Size Scaling}

The clock model is defined as a classical spin chain of length $L$
with dynamical variables $\vec{s_n}$ fluctuating among the $q$ roots
of unity
\be
\vec{s_n} = \left( \cos {2\pi p_n \over q} , \sin {2\pi p_n \over q}
           \right); \qquad p_n = 0,1,\ldots,q-1
\ee
and interacting through the Hamiltonian
\be
\beta {\cal H} = E L =
    - \sum_n \left( J \cos {2\pi \over q}(p_n - p_{n+1}) +
          i \,\epsilon\, \sin {2\pi \over q}(p_n - p_{n+1}) \right)
\label{ham}
\ee
where $J$ and $\epsilon$ are real coupling constants (the
$\beta$ factor has been absorbed
absorbed in the definition of these couplings).  The
model can also be seen as a chiral $q$-state Potts model and also
describes a $\IZ_q$ gauge theory in a cylindric lattice
\cite{wheather}. Due to the presence of the imaginary terms in
the Hamiltonian \reff{ham}, the model undergoes first order phase
transitions \cite{asorey-esteve}.
The behavior of the renormalization group for the $q=3$ and $q=4$
agrees with the discontinuity fixed point scenario
and  presents   pathologies similar to the
Griffiths-Pearce pathologies \cite{clocks}.
Although the models are very
peculiar because of the complex character of the Hamiltonian, some
of our conclusions are also relevant for models with real-valued
interactions with first order transitions of infinite correlation
length.

In this note we will concentrate on  the analysis of the behavior of
the finite size corrections in the $q \rightarrow \infty$ limit.
In that case the
model corresponds to a $U(1)$ gauge theory with a $\theta$-term in
a two dimensional cylindrical lattice \cite{wheather,rusakov}.
The eigenvalues of the transfer matrix are \cite{asorey-esteve}
\be
\label{eigenvalues}
\lambda_k = \left\{  \begin{array}{ll}
       2\pi \left( {J+\epsilon \over J-\epsilon} \right)^{k/2}
       I_{|k|}(\sqrt{J^2 - \epsilon^2}) & |\epsilon| \leq J \\
       2\pi \left( {J+\epsilon \over \epsilon-J} \right)^{k/2}
       J_{|k|}(\sqrt{\epsilon^2 - J^2}) & |\epsilon| > J
       \end{array} \right.
\ee
where $I_k$ and $J_k$ are the Bessel functions of integer order.
If we  consider  periodic boundary conditions
only the leading eigenvalue of the transfer matrix is relevant
for the  thermodynamic limit.
The phase transitions occur at the points $(J,\epsilon)$ of the
parameter space where the leading eigenvalue is degenerate. The
energy density is discontinuous at these points, thus the
transitions are first order, although they have a divergent
correlation length $\xi=\infty$. This unusual feature is due to the
complex character of the action. Furthermore, the critical index
$\nu$ associated to the divergency of the correlation length is $\nu
= 1$, in agreement with the discontinuity fixed point picture for
first order phase transitions \cite{dfp}.

For simplicity we will only consider phase transition points with $J =
\epsilon$, but the results can easily be
generalized for all transition points. In this case the
eigenvalues of the transfer matrix \reff{eigenvalues} read
\be
\label{eigenvalues2}
\lambda_k = \left\{  \begin{array}{ll}
       2\pi {J^k / k!} & k \geq 0 \\
       0     & k < 0
       \end{array} \right.
\ee
which implies the existence of first order phase transitions at the
points
\be
J_c^{(k)} = k+1 \ ; \qquad k=0,1,2,\ldots
\ee
The mean energy density is given by
\be
\label{energy_density}
u(J,L) = \langle E \rangle_L = {1 \over L}
         {\partial \over \partial J}  \log Z(J,L),
\ee
where
\be
Z(J,L) = \sum_{ \{ p_n \} } e^{-EL} = \sum_{k \geq 0}
                         (\lambda_k)^L
\ee
is the partition function of a periodic chain of length $L$.
In the thermodynamic limit ($L \rightarrow \infty$) we recover the
internal energy density
$u(J)=\langle E \rangle = \lim_{L \rightarrow \infty}\langle E
\rangle_L $.

The spin correlation function
\be
\label{correlation}
 \left\langle \vec{s_0} \cdot \vec{s_m} \right\rangle_L =
          {1 \over 2} \frac{   \sum_{k \geq 0}
          \left( {J^k \over k!} \right)^L \left\{
     \left( {J \over k+1} \right)^m + \left( {k \over J} \right)^m
                                         \right\}  }
     {\sum_{k \geq 0}  \left( {J^k \over k!} \right)^L }.
\ee
is dominated in that limit by the leading eigenvalue
\be
\left\langle \vec{s_0} \cdot \vec{s_m} \right\rangle =
\lim_{L \rightarrow \infty}\left\langle \vec{s_0} \cdot \vec{s_m}
\right\rangle_L
       = {1 \over 2} \left\{ \left( {J \over 1 + [J]} \right)^m +
       \left( {[J] \over J} \right)^m \right\},
\label{correlation2}
\ee
where $[J]$ denotes the integer part of $J$. Near the transition point
the correlation function  $\left\langle \vec{s_0} \cdot \vec{s_m}
\right\rangle $ has an asymptotic exponentially decaying behavior
\be
\label{massgap}
\left\langle \vec{s_0} \cdot \vec{s_m} \right\rangle \sim
 A e^{-\mu(J) m}
\ee
as $m$ goes to $\infty$, with  mass gap  $\mu(J)$  given by
\be
   \mu(J) = \frac{|J - J_c^{(k)}|}{J_c^{(k)}}
\ee
for $|J - J_c^{(k)}| \ll J_c^{(k)}$. Thus, the correlation length
$\xi(J)=\mu(J)^{-1}$ diverges as $J$ approaches the critical point
with a critical exponent $\nu = 1$.

On the other hand, one can expand Eq.~\reff{correlation2} around a
transition point. Using the analogue of Eq.~\reff{massgap} for finite
$L$ we get an expansion for the mass gap in powers of
$\eta = J-J_c^{(k)}$ and $N$ plus some negligeable
exponential terms. In the particular case  $J_c^{(0)} = 1$  the result is
\begin{eqnarray}
\mu^2(J=1+\eta,L) &=& 2^{-L/2} \log^2 2 + \eta e^{-L/2}
                  \left[ L \log^2 2 - 2 \log 2 \right] + \nonumber \\
 & + &       \eta^2 \left[ 1 + {\cal O}(2^{-L/2}) \right] +
       {\cal O}(\eta^3,2^{-L}).
\end{eqnarray}
The minimum value of the mass gap
\be
\mu^2_{min} = 2^{-L/2} \log^2 2 + {\cal O}(2^{-L})
\ee
is reached at the points
\be
J( \mu^2_{min} ) = 1 - {L \over 2} e^{-L/2} \log^2 2
   \left( 1 - {2 \over L \log 2 } \right) + {\cal O}(2^{-L}).
\ee
which is in agreement with the results obtained above for the
thermodynamic limit $L \rightarrow \infty$. In Fig.~1 we have plotted
the function $\mu(J,L)$ near $J_c^{(0)}$ for several values of $L$.

An interesting property of the model is that the
derivative  of the mass gap with respect to $L$ is negative near a
transition point. In particular, in a neighborhood of
$J_c^{(0)}$ we have that
\be
{d\mu^2 \over dL} = - {L \over 2} 2^{-L/2} \log^2 2 \left\{
       \log 2 + {\cal O}(\eta) {1 \over L} \right\} < 0
\ee
This means that the curves $\mu(J,L)$ do not intersect each other
near the transition point (See Fig.~1). This behavior does not agrees
with the picture advocated in Ref.~\cite{okawa} to discriminate between
first and second order phase transition. In this model the  curves
$\mu(J,L)$ which describe the finite size approximation to the mass gap
do not cross each other in the vicinity of first order phase
transition points. Thus, the absence of crossing of these curves
cannot be taken as a  signal of second order phase transition for any
statistical-mechanical model. In fact, it is likely associated to the
infinite correlation length of the system irrespectively of the first
or second order nature of the phase transition. In particular,
it can be shown that the
same behavior arises for all spin models whose lowest  energy
levels are  degenerated with the only condition that the spin variables
must have a non--vanishing correlation between some of those
degenerated eigenstates.

In the thermodynamic limit the specific heat is a non-positive and
discontinuous function of the coupling constant $J$
\be
     CV(J)   =  - [J]
\ee
The jumps on the specific heat are associated to the different phase
transitions.

Near the phase transition points the finite size corrections
\be
\label{cv_def}
CV(J,L) = J^2 L \left( \langle E^2 \rangle_L -
                       \langle E   \rangle_L^2   \right)
\ee
can be  expanded in  terms
of powers in $\eta=J-J_c^{(k)}$, ${1/ L}$ and exponentially
decaying terms of the form $\exp (-a L)$. If we neglect the
exponential terms $CV(J_c^{(k)}+\eta,L)$ reduces to a polynomial in
$\eta$ with $L$-dependent coefficients. In this way we obtain the
coupling $J(CV_{max})$ where the specific heat $CV$ reaches a maximum
value  $CV_{max}$. The
results for $J_c^{(0)} = 1$ are given by
\begin{subeqnarray}
J(CV_{max}) &=& 1 - {2 \over L^2} + {\cal O}(L^{-4}) \\
    CV_{max} &=& {L \over 4} - {1 \over 2} + {1 \over 4L} +
                 {\cal O}(2^{-L})
\end{subeqnarray}

The asymptotic behaviors of those quantities are similar to the
ones given by BK for the Potts model. However, we find a clear
difference in $J(CV_{max})$. The leading correction to $J(CV_{max})$
is ${\cal O}(L^{-2})$, instead of ${\cal O}(L^{-1})$. In the
discontinuity fixed point interpretation this means that the shift
of the $J(CV_{max})-J_c$ does not scale as $L^{-1/\nu}$.

Nevertheless, the specific heat behaves as predicted by BK and in
agreement with a critical exponent $\alpha = 1$.
Finally, we note that the value of $CV$ at the transition point picks up
only exponentially small corrections, as observed in Ref.
\cite{billoire10}
\be
CV(1,L) = {L \over 4} - {1 \over 2} + {\cal O}(2^{-L})
\ee
This means that the asymptotic regime is reached  much faster at
the thermodynamic transition point $J=1$ than at the finite size
maximal point $J(CV_{max})$ (See Fig.~2).

Now we examine the behavior of the  $U4(J)$ ratio of cumulants in a
finite chain,
\be
U4(J,L) = \frac{ \langle ( E - \langle E \rangle_L )^4 \rangle_L }
               { \langle ( E - \langle E \rangle_L )^2  \rangle_L^2 }.
\ee
In the thermodynamic limit it takes the following form
\be
U4(J) = \left\{ \begin{array}{lll}
              \infty &  J < 1        &                  \\
                3    &  J > 1        &  J \neq J_c^{(k)} \\
                1    &  J = J_c^{(k)} &  k=0,1,2,\ldots
              \end{array}  \right.
\ee
The divergency of $U4(J)$ for $J \in [0,1)$ is due to the fact
that the density energy is constant in this interval. Notice that
the value $U4(J_c^{(k)}) = 1$ agrees with the expected result for
first order transitions.

Repeating  the  analysis  of finite size approximation for the $U4$
ratio of cumulants we get the following values for its minimum value
\be
U4_{min} = 1 - {8 \over L} - {4 \over L^2} +
               {\cal O}(L^{-3})
\ee
and its position
\be
 J(U4_{min}) = 1 - {4 \over L^2} + {\cal O}(L^{-3})
\ee
in the vicinity of the transition point $J= J_c^{(0)}$.
Notice that $U4< 1$ and this property  can  never
arise for systems with  real-valued interactions, but in this case
it is gernerated by the
complex character of the action. The leading correction to the
position of the minimum is again of order ${\cal O}(L^{-2})$.

We now repeat the same analysis for the Binder--Landau cumulant,
\be
BL(J,L) = {1 \over 3} \left( 1 - \frac{\langle E^4 \rangle_L}
          {\langle E^2 \rangle^2_L}  \right),
\ee
which in the thermodynamic limit  takes the form
\be
BL(J) = \left\{ \begin{array}{cll}
              -\infty &  J < 1        &                  \\
                0    &  J > 1        &  J \neq J_c^{(k)} \\
               -{1 \over 12}       { (1+2k)^2 \over k^2(1+k)^2 }
                     &  J = J_c^{(k)} &  k=0,1,2,\ldots
              \end{array}  \right.
\ee
The explicit value for $BL(J_c^{(k)})$ can be obtained following
Ref.~\cite{billoirePRB} and using the fact that the internal energy at
the transition points jumps from $k/(k+1)$ to 1.

The maximum value of the cumulant and its position near $J_c^{(1)} = 2$
are given by (see Fig. 3)
\begin{subeqnarray}
BL_{min} &=& - {3 \over 16} - { 35 \over 32 L}
             - {751   \over  1728 L^2} +  {\cal O}(L^{-3})\\
  J(BL_{min}) &=& 2 - {2 \log 4\over L} +
                   {23+ 9 \log^2 4 \over 9 L^2} + {\cal O}(L^{-3})
\end{subeqnarray}

The leading correction to $J(BL_{min})$ is, amazingly, of
order ${\cal O}(L^{-1})$ which is in full agreement with the BK
ans\"atz, but slightly different of the  results for the other
cumulants. In
this model only the Binder-Landau cumulant scales as $L^{-1/\nu}$.

\section{Conclusions}

In summary, the BK ans\"atz is
correctly reproduced in this simple model. The values
of $CV_{max}/L^d$, $BL_{min}$ and $U4_{min}$ are in agreement with
those associated to first order transition points. The complex
character of the action only generates some minor modifications: the
specific heat is non-positive and the  $U4$ ratio of cumulants is
smaller than one for some values of $J$.

The scaling of the position of the minimum of $BL(J,L)$ is also
in agreement with the BK ans\"atz: the leading correction behaves as
$L^{-1/\nu}=L^{-d}$. However, the leading corrections for the extremal
points of $U4(J,L)$ and $CV(J,L)$ are of order $L^{-2}$.
This behavior does not implies the failure of the standard
finite-size scaling \cite{barber}, but simply means that the
leading correction is
not {\em universal}, because it might depend on the quantity
considered.

Finally,  the finite size approximations to the mass
gap in this model do not behave as for the first order phase
transitions of
 Ising and Potts models in two dimensions. The
mass gap curves $\mu(J,L)$ obtained for different sizes of the
 system  do not
cross each other in the vicinity of first order transition points.
Therefore this property cannot be used as an indicator of the
 order of the phase transition
\cite{okawa}.
It rather indicates the existence of an infinity length correlation
in the system.

\subsection*{Acknowledgements}
We thank Juan Jes\'us Ruiz-Lorenzo for helpful discussions.
We acknowledge the financial support of the CICyT, through grants
AEN90-29 and AEN90-30.

\newpage
\section*{Figure Captions}

\noindent {\bf Figure 1:}
Dependence of the mass gap  $\mu(J,L)$ on the
coupling $J$ in a neighbourhood of $J_c^{(0)}=1$ for different values
of the chain length: $L$ = 20, 25 and 30.
For $L \geq 40$ the curves overlap
almost perfectly with the thermodynamic result (thicker line).

\vspace*{0.5cm}

\noindent {\bf Figure 2:}
Behavior  of the specific heat  function $CV(J,L)$ in the same
region as in Fig.~1. We plot the curves for $L=20,40$ and 80
together with the thermodynamic limit (thicker line).

\vspace*{0.5cm}

\noindent {\bf Figure 3:}
Finite size behavior of the Binder-Landau cumulant $BL(J,L)$ near
$J_c^{(1)}=2$. The symbol $\bullet$ marks the thermodynamic limit at
$J_c^{(1)}=2$.

\end{document}